\documentclass[12pt,a4paper]{article}
\usepackage{amsfonts,graphicx,bezier}
\usepackage{amsfonts,amssymb, float,ctable}
\usepackage{graphicx,enumerate}
\newtheorem{thm}{Theorem}[subsection]
\newtheorem{defn}{Definition}[section]
\newtheorem{prop}{Proposition}[subsection]
\newtheorem{lem}{Lemma}[subsection]
\title{\large {\bf A Mathematical Analysis and Evaluation of Varroa mite Control Methods for a Mite-infested Honey bee Colony}} 
 \vspace{1.8cm}
\author{{\footnotesize {\bf  A. Ssenoga$^{a}$, H. Ddumba$^{b,*}$ \& J.Y.T. Mugisha$^{b,*}$}}\\ 
{\footnotesize $^b$Department of Mathematics, Makerere University, P.O.Box 7062, Kampala-Uganda.}\\{\footnotesize Email ($^*$Correspondences): hassan.ddumba@mak.ac.ug, joseph.mugisha@mak.ac.ug.}}
\date{}
\begin{document}
\maketitle
\vspace{0.01cm}
{\footnotesize  
\paragraph\ {\bf Abstract:} In this paper, we present a mathematical model for the interaction between honey bees and mites. The dynamics of a mite-infested honey bee colony and the evaluation of the commonly used mite-control strategies (traditional, mechanical and chemical) are studied. The mite-free and mite reproduction numbers $R_{h}$ and $R_{m}$ respectively, are derived using the next generation operator approach. The mathematical analysis of the model reveals that in the absence of mites, the colony survives if $R_{h}>1$ otherwise it goes extinct if $R_{h}<1.$ Stability and sensitivity analyses of the model reveal that the egg laying rate of the queen bee is key in regulating mite reproduction. Adult bee grooming and hygienic behavior of worker bees have also been shown to play a vital role in reducing parasitism. Using the Volterra-Lyapunov stable matrix approach, the mite-infested equilibrium is confirmed to be globally asymptotically stable when $R_{h}>1$ and $R_{m}>1.$ It is also shown that varroa mite control strategies that focus on limiting mite-reproduction such as caging the queen bee and using a young queen bee quickly reduces $R_{m}$ to a value less than unity compared to those that are intended to kill the mites.
\vspace{0.3cm}}
\section{Introduction}
\paragraph\ The loss of honey bee colonies and the reduction in colony productivity due to Varroa mites are major concerns by bee keepers around the world (Kang {\emph{et al}}., 2016). Mites feed on the bee brood and adult bees (Messan \emph{et al}., 2012; Kang \emph{et al}., 2016). They live externally on bee bodies as phoretic mites (PM) and inside brood cells as reproductive mites (RM) (Messan \emph{et al}., 2012). Varroa-mites suck nutrients from honeybees resulting in reduced vigour, shortened life span of bees and eventual collapse of colonies (Di Prisco \emph{et al.,} 2016). Varroa-mite infested bees have a shorter lifespan than the un-infested bees. The mite-infested bees are also unable to maintain their roles. This failure of bees to maintain their roles, negatively affects egg-production. Heavy mite-infestation causes scattered brood, crippled and crawling honey bees, impaired flight performance and a lower rate of return to the colony (Di Prisco \emph{et al.,} 2016). Hence, heavy mite-infestation negatively affects all bee activities leading to a reduction in colony strength and productivity (colony performance). 
\paragraph\ Varroa mites have spread to every part of the world except Australia (Boncristiani \emph{et al}., $2021)$. This is the reason why many bee-keepers around the world are carrying out management practices to reduce the mites in their colonies rather than carrying out preventive measures. These commonly used management practices include; traditional, mechanical and chemical control methods (Cameron and James, 2021). Some of the traditional control methods include; use of hygienic bees that remove mite-infested brood, caging the queen temporarily to cause a brood break thus limiting mite reproduction, using a young queen to produce more worker brood and less drone brood, comb culling and sterilization of hive equipments. Mechanical control methods include; use of screened bottom boards to prevent mite invasion into brood cells and drone brood trapping to limit mite reproduction opportunities. The chemical control methods include; spraying the bees with both hard and soft chemicals. In this case, bee-keepers are advised to use soft organic chemicals such as formic acid because they are harmless to the bees (Marco and Giovanni, $2018$; Ziyad \emph{et al}., $2021$). They also penetrate inside the pupae cells to kill the reproductive mites (Fries \emph{et al}., $1991$).
\paragraph\ Various Mathematical models have been formulated to study the effect of mites on honey bee populations (Kang \emph{et al}., 2016). In these studies, sensitivity analyses on model parameters have been carried out to investigate the most sensitive parameters. In these studies, sensitivity analyses were used to identify key parameters and subsequently effective mite control strategies were suggested. Numerical simulation results were used to show the effect of different parameters on colony performance.
\paragraph\ Denes and Mahmoud (2019) formulated and analysed a mathematical model for the dynamics of a honey bee colony affected by mites. They divided the bees into compartments depending on their infestation and infection status instead of creating separate mite compartments. Their study shows that it is possible to eliminate the mites from the colony if the mite reproduction number is less than one. They also suggest different mite-control methods basing on the basic reproduction numbers.
\paragraph\ Bernadi and Venturino (2016) used a mathematical model to describe how the mites affect the virus epidemiology on adult bees. The mite birth rate, the bees' disease mortality rate and horizontal virus transmission rate were found to be the most influential parameters in their study.
\paragraph\ Kang \emph{et al.,} (2016) proposed a general bee-mite-virus model to determine the effects of parasitism and virus infections on bees. They considered proportions of brood and adult bees in order to determine the effect of parasitism on bees. The results of their model analysis show that varroa mite control strategies must be focussed on keeping mite levels and virus transmission rates as low as possible. Brood rearing resumes in early spring prompting mite population numbers and their growth rates to increase. Therefore, Kang \emph{et al.,} (2016) suggest that mite spraying (management) should be intensified in early spring.
\paragraph\ Torres and Torres (2020) used differential equation models to track each day in the life of a bee and uses different survival rates for each of the different bee castes. They performed model simulations to reveal that colony survival is sensitive to adult bee grooming rate and reproductive rate of mites. They also suggest bee management processes such as drone brood removal and selective breeding in order to improve colony health. In the above studies, the authors recommended to the bee keepers the possible mite control strategies. However, subsequent studies have not mathematically analysed and scientifically evaluated these commonly used mite control methods.    
\paragraph\ This study makes use of a host-parasite modelling approach by Kang \emph{et al}., (2016). Since mites reproduce inside the pupae cells, in this study an explicit pupae population is modelled in order to correctly estimate the mite population. Because the reduction in the hatching rate causes a delayed recruitment of new bees, the hatching period of the pupae is also determined. A long hatching period provides ample time for mite maturation inside the pupae cells, thus increasing mite population growth.
\paragraph\ An investigation on how parasitism through mite-feeding activities affects the numbers of pupae cells and adult bees is carried out. The long term behavior of a mite-infested colony is studied via local and global stability of equilibrium points. Using sensitivity and numerical analyses on model parameters, a mathematical evaluation of mite control methods currently practiced by bee keepers is carried out.
     
\section{Methods}
\subsection{Model Description and Formulation}
\paragraph\ Varroa mites need bees to survive because they either live on adult bee bodies (as phoretic mites) or inside brood cells (as reproductive mites). This therefore, calls for a scientific study of the interaction between bees and mites. A honey bee colony is categorised into three compartments namely: pupae cells $(x),$ adult honey bees $(y)$  and mites $(z).$ The pupae population is modelled because the mites reproduce inside the pupae cells (Messan \emph{et al}., 2012; Kang \emph{et al}., 2016). The mites feed on the immature bees inside the pupae cells which facilitates their reproduction. The model description, together with assumptions, parameters and variables is as follows: The queen lays eggs at a maximum rate $\Lambda$ per day which metamorphoses into larvae, pupae and then into adult workers. The brood (egg,larvae and pupae) require care from adult bees so as to hatch into adult bees. The adult bees in this case incubate the pupae cells to facilitate the hatching process. The pupae population is multiplied by the fraction $\frac{y}{K+y}$ that determines the number of pupae cells that hatch into adult bees (Khoury \emph{et al.,} 2011). The constant $K,$ which  is the half saturation constant, which refers to the number of workers required for 50\% of pupae cells to grow into adult bees.
\paragraph\ The pupae cells $(x)$ are assumed to be recruited at a rate directly proportional to the egg laying rate, $\Lambda$ of the queen bee. Therefore, the number of pupae cells recruited at any time $t$ is given by $\frac{\Lambda y}{K+y}$ (Khoury \emph{et al}., 2011).  The limiting behavior $\lim\limits_{y\rightarrow\infty}\frac{\Lambda y}{K+y}=\Lambda,$
implies that a sufficiently large number of workers is required for the efficient egg-production and brood rearing. The pupae hatches into adult workers at a rate $\sigma$ per day but also dies naturally at a rate $m$ per day. 
\paragraph\ Considering the proportions of phoretic mites (PM) and reproductive mites (RM) to be $\epsilon$ and $(1-\epsilon),$ respectively, then $\epsilon z(t)$ and $(1-\epsilon)z(t)$ are the populations of PM and RM at any time t, respectively. The $RM$ feeds on the immature bees at pupae stage and reduces its lifespan at a rate $\hat\alpha.$ The RM convert the nutrients obtained from the immature bee at a rate $c$ in order to support mite reproduction. The mite population dies due to hygienic behavior at a rate $h,$ dies due to grooming at a rate $g$ and dies naturally at a rate $d_{m}$ whereas the adult bees die naturally at a rate $\mu.$
\paragraph\ The following assumptions are applicable to this study: The pupae population is recruited at a rate directly proportional to the egg laying rate of the queen bee. Parasitism by mites does not reduce the life span and does not cause death of adult bees (Sumpter and Martin, 2004). The adult drones do not participate in colony work and therefore not considered in this study. The hive and forager bees are taken as adult bees because both categories support egg-production and brood rearing through division of labour. The queen bee is not infested by mites, implying that its egg-laying rate is not affected by its health but by the number of workers that rear the brood. The outlined model description leads to the compartmental diagram in Figure $1$ while the model parameters are described in Table $1.$
\begin{figure}[H]
%\begin{minipage}{0.3\textwidth}
  \centering
  \includegraphics[width=10cm]{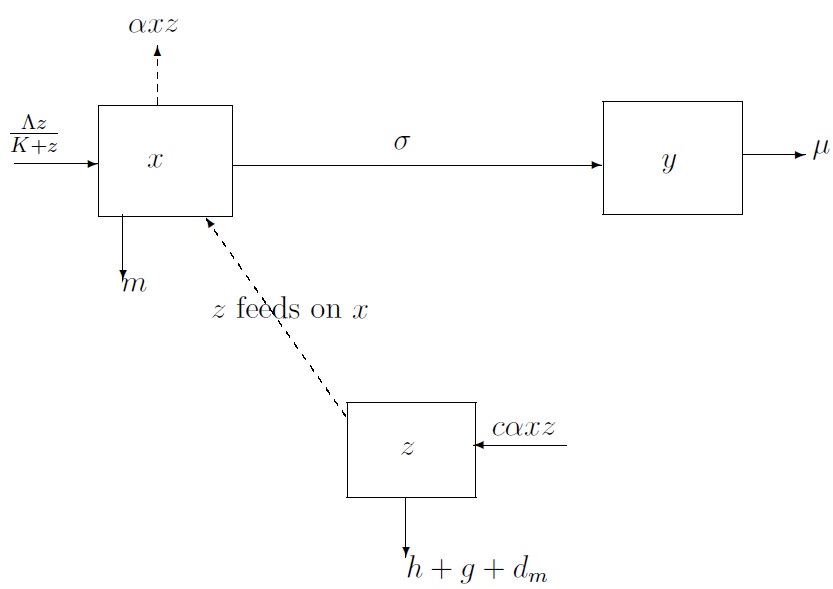}
%\end{minipage}\hfil
\caption{ \it A Flow diagram showing the population dynamics of honeybees infested with mites.} 
\end{figure}

%\begin{figure}%[h]
%\centerline {\includegraphics[scale=0.5]{mn.jpg}}
%\caption{ \it A Flow diagram showing the population dynamics of honeybees infested with mites.} 
%\end{figure}
\begin{table}[h!]
\caption{Model parameters and their estimated values}
\centering
\begin{tabular}{lllll}
  \toprule
  Parameter& description&value&Units&Source\\
  \midrule
$\Lambda$ & maximum egg-laying rate of the queen bee&1500&$day^{-1}$&Khoury \emph{et al}. 2011\\
$\hat\alpha$ & parasitism rate on pupae &[0,1]&-&Torres $\&$ Torres, 2020  \\
$\mu$ & adult bee natural mortality rate&0.04& $day^{-1}$&Khoury \emph{et al}., 2011\\
$d_{m}$ &  natural mite mortality rate&0.006& $day^{-1}$&Messan \emph{et al}., 2012\\
$m$ & pupae natural mortality rate&0.001&  $day^{-1}$&Torres $\&$ Torres, 2020\\
$h$& rate of hygienic behaviour &0.08&$day^{-1}$ & Santos \emph{et al}., 2016\\
$g$& rate of adult bee grooming &$0.05$&$day^{-1}$ &Torres $\&$ Torres, 2020\\
$c$ & nutrient conversion rate by mites&[0,1]&-&Bernadi \& Venturino,2016 \\
$\epsilon$ & proportion of phoretic mites in the colony&[0,1]&-&proportion\\
$K$ & half-saturation constant&5000&bees&Khoury \emph{et al}., 2011\\
$\sigma$& emergence rate of adult bees&$\frac{1}{12}=0.083$&$day^{-1}$&Schmickl $\&$ Karsai, 2016\\
\bottomrule
\end{tabular}
\end{table}
 \paragraph\ The description above results into the following system of model equations:
\begin{eqnarray}
\frac{dx}{dt}&=&\frac{\Lambda y}{K+y}-(\sigma+m)x-\alpha xz\nonumber\\
\frac{dy}{dt}&=&\sigma x-\mu y\\
\frac{dz}{dt}&=&c\alpha xz-\mu_{3}z,\nonumber
\end{eqnarray}
where $\alpha=\hat\alpha(1-\epsilon),$ is the rate at which reproductive mites attack the pupae cells and $\mu_{3}=d_{m}+g+h,$ is the total mite mortality rate.
\subsection{Model Analysis}
\subsubsection{On Boundedness and Positivity of Solutions}

\paragraph\ Let the initial population of honey bees and mites be, $\{x(0)>0,y(0)>0,z(0)\geq0\}\in \omega.$ Then the solutions $\{x(t),y(t),z(t)\}\in \omega$ of System (1) are positive for all $t>0.$
%\textbf{Proof}:
If all parameters are positive in System (1), then:
$\frac{dx}{dt}|_{x=0} =\frac{\Lambda y}{K+y}>0,$ $\frac{dy}{dt}|_{y=0} =\sigma x>0$ and $\frac{dz}{dt}|_{z=0} =0.$ Thus, $\omega=\{(x,y,z):x>0,y>0,z\geq0\}$ is a positively invariant set that attracts all model trajectories (Thieme, 2003). It is important to establish whether the model is epidemiologically well posed. In this case, we study the invariant region in which solutions of System (1) are biologically meaningful.
\paragraph\ Let $T=c(y+x)+z$ as in Kang \emph{et al}. 2016, then:
\begin{eqnarray*}
\frac{dT}{dt}&=&c\frac{dx}{dt}+c\frac{dy}{dt}+\frac{dz}{dt}\\
&=&c\Lambda\frac{y}{K+y}-cmx-c\mu y-\mu_{3}z
\end{eqnarray*}
The limiting behavior, $\lim\limits_{y\rightarrow\infty}(\frac{\Lambda y}{K+y})=
\lim\limits_{K\rightarrow 0}(\frac{\Lambda y}{K+y})= \Lambda,$
means that a sufficiently large number of workers or a small number of efficient workers attending to the queen bee is required for the maximum egg-laying rate to be attained. Bee keepers are therefore advised to increase the strength of their colonies so that sufficient amounts of nectar and large quantities of pollen can be brought to the hive. This will promote the well being of the queen enabling it to lay as many eggs as it can. With these assumptions, it follows that;
\begin{eqnarray}
\frac{dT}{dt}&\leq&c\Lambda-cmx-c\mu y-\mu_{3}z\nonumber\\ \\
&\leq&c\Lambda-\mbox{min}\{m,\mu,\mu_{3}\}(c(x+y)+z).\nonumber
\end{eqnarray}
Solving inequality $(2)$ yields;

$$T(t)\leq\frac{c\Lambda}{\gamma}+(T(0)-\frac{c\Lambda}{\gamma})e^{-\gamma t},$$                 
where $T(0)>0$ and $\gamma=\mbox{min}\{m,\mu,\mu_{3}\}.$ Therefore as $t\rightarrow\infty,$ $T(t)\leq\frac{c\Lambda}{\gamma}$ meaning that every solution that originates inside $\omega$ remains in $\omega$ for all values of $t.$ Hence, $X$ is feasible and positively invariant. This implies that System (1) is mathematically and epidemiologically well posed. Now, assuming a maximum egg-laying rate, $\Lambda$ by the queen bee and a mite-free colony $(i.e., \alpha=0)$ the upper bound of the pupae cells is found as follows:
\begin{eqnarray}
\frac{dx}{dt}&=&\frac{\Lambda y}{K+y}-(\sigma+m)x-\alpha xz\nonumber\\ \\
&\leq&\frac{\Lambda y}{K+y}-(\sigma+m)x\leq\Lambda-(\sigma+m)x.\nonumber
\end{eqnarray}
\paragraph\ Therefore $\lim\limits_{t\rightarrow\infty}\mbox{sup}\, x(t)\leq\frac{\Lambda}{\sigma+m},$ which implies that the pupae population is bounded. This bound is directly proportional to the egg laying rate $(\Lambda)$ and inversely proportional to the total depletion rate of the pupae cells through pupae mortality and adult bee emergence. The upper bound of the adult bees is found by assuming a maximum egg laying rate so that;
\begin{eqnarray}
\frac{dy}{dt}=\sigma{x}-\mu{y}\leq \Lambda \frac{\sigma}{\sigma+m}-\mu{y}
\end{eqnarray}
\paragraph\  This implies that $\lim\limits_{t\rightarrow\infty}\mbox{sup}\, y(t)\leq\Lambda\frac{\sigma}{\mu(\sigma+m)}$ which shows that the adult bee bound is directly  proportional to the product of the egg laying rate, $(\Lambda)$ the proportion of adult bees emerging out of pupae, $\frac{\sigma}{m+\sigma}$ and inversely proportional to the adult bee mortality rate $(\mu).$  Bee keepers should carry out practices that enables the queen to lay as many eggs as it can. These include increasing colony strength which enables the colony to fetch enough food resources to the hive thus, increasing colony performance. The adult bees also live longer when they are fed well. This enables them to effectively and efficiently do colony duties such as brood rearing which is key to increasing colony strength. In order to achieve this, Bee keepers should monitor their colonies to avoid attacks from mites, viruses and termites that may cause high bee mortality.
\subsubsection{Stability Analysis of Equilibrium Points}
\paragraph\ Model (1) admits three equilibria; the extinction equilibrium $E_{0}(0,0,0)$, the mite-free equilibrium $E_{f}(\frac{\Lambda}{m+\sigma}-\frac{K\mu}{\sigma},\frac{\Lambda\sigma}{\mu(m+\sigma)}-K,0)$ and the mite-infested equilibrium $E_{i}(x^{*},y^{*},z^{*})$ where
\begin{eqnarray}
  x^{*}=\frac{\mu_{3}}{c\alpha},~~~~y^{*}=\frac{\sigma\mu_{3}}{c\alpha\mu},~~~~z^{*}=\frac{1}{\alpha x^{*}}[\frac{\Lambda y^{*}}{K+y^{*}}-(\sigma+m)x^{*}]
\end{eqnarray}
\textbf{The Basic Reproduction Numbers, $R_{h}$ and $R_{m}$}
\paragraph\ Two basic reproduction numbers $R_{h}$ and $R_{m}$ are considered to be of great significance for the model. $R_{h}$ is the mite-free (demographic) reproduction number for the model and it provides conditions under which a honey bee colony can survive or avoid getting extinct under demographic parameters. The mite-free equilibrium $E_{f}$ of Model (1) exists if $\frac{\Lambda\sigma}{K\mu(m+\sigma)}>1$ with the left hand side of the inequality providing the basic reproduction ratio;
\begin{eqnarray}
R_{h}=\frac{\Lambda\sigma}{K\mu(m+\sigma)}
\end{eqnarray}
\paragraph\  The term $\frac{\Lambda\sigma}{K(m+\sigma)}$ represents the number of adult bees that emerge from the pupae per adult bee that takes care of the brood. The term $\frac{1}{\mu}$ defines the average life span of the adult bee that takes care of the brood. Therefore, $R_{h}$ is the number of adult bees emerging from the pupae when one adult bee takes care of the brood in its entire life span as an adult bee. On the other hand, $R_{m}$ which is the mite reproduction number, defines the number of new mites born to a single mother mite during its entire lifespan when introduced in a completely mite-free colony. $R_{m}$ determines whether the mites will spread or will be eradicated if a single mite is  introduced in a completely susceptible bee population. Using the next generation matrix method by van den Driesche and Watmough (2002) and referring to the third equation of Model $(1),$ the Jacobian matrix of new infestation (F) and that of all transition terms (V) evaluated at the mite-free equilibrium, $E_{f}$ are given by  F=c$\alpha x^{*}$ and V=$\mu_{3}$ so that;
\begin{eqnarray}
R_{m}=\rho(FV^{-1})=\frac{c\alpha x^{*}}{\mu_{3}}=\frac{c\alpha}{\mu_{3}}[\frac{\Lambda}{\sigma+m}-\frac{K\mu}{\sigma}]
\end{eqnarray}
\paragraph\ On the existence of the equilibrium points; the extinction equilibrium $E_{0}$  exists for all parameter choices, the mite-free equilibrium $E_{f}$ exists if $\Lambda\frac{\sigma}{m+\sigma_{1}}>K\mu$ which on re-arrangement is equivalent to $R_{h}>1.$  The mite-infested equilibrium $(E_{i})$ exists if $x^{*}>0,$ $y^{*}>0$  and
$z^{*}>0.$  $z^{*}>0$ iff $\frac{\Lambda y^{*}}{K+y^{*}}>(\sigma+m)x^{*}.$ Substituting for $x^{*}$ and $y^{*}$ from $(5)$ into the inequality simplifies to$\frac{\Lambda}{m+\sigma}-\frac{K\mu}{\sigma}>\frac{\mu_{3}}{c\alpha}$ which on re-arrangement holds if $R_{m}>1$ when $R_{h}>1.$ This shows that the mite-infested equilibrium $E_{i}$ exists when $R_{m}>1$ and $R_{h}>1.$
\subsubsection{Local stability analysis of $E_{0},$ $E_{f}$ and $E_{i}$}
\paragraph\ To study the long term behavior of Model (1), we investigate the stability of the three equilibria. The Jacobian matrix (J) of Model $(1)$ is computed and  defined as;
\begin{eqnarray}
 J=\left[\begin{array}{ccccc}-(m+\sigma+\alpha z^{*})&\frac{\Lambda K}{(K+y^{*})^{2}}&-\alpha x^{*}\\\sigma &-\mu & 0\\ c\alpha z^{*}&0&c\alpha x^{*}-\mu_{3} \end{array}\right]
\end{eqnarray}
At $E_{0},$ $J|_{E_{0}}$ has a negative eigenvalue, $\lambda_{1}=-\mu_{3}.$ By descartes rule of signs, the remaining eigenvalue is negative if $\frac{\Lambda\sigma}{K\mu(m+\sigma)}<1$ implying that $ R_{h}<1$ which provides the stability condition for $E_{0}$.  This condition shows that bee keepers should carry out practices that increase; the egg-laying rate $(\Lambda)$, the proportion of pupae that become adult bees $(\frac{\sigma}{m+\sigma}),$ but reduce; adult bee mortality rate $(\mu)$ and adult bees that care for the pupae $(K)$. Therefore, bee-keepers need strong colonies that would forage sufficient amounts of nectar and pollen. This facilitates the queen's well being which increases egg production thus, avoiding colony extinction. Strong colonies also effectively incubate a large proportion of the pupae cells so that there is no delay in the birth of new bees which increases bee population.
{\thm The mite-free equilibrium $E_{f}$ is locally asymptotically stable if $R_{m}<1$ and $R_{h}>1.$}\\
{\bf Proof:} The system settles to a mite free-equilibrium if the characteristic equation of $J|_{E_{f}}$ has only negative eigenvalues. Algebraic computation yields $\lambda_{1}=\mu_{3}(R_{m}-1)<0$ if $R_{m}<1.$ The remaining two eigenvalues $\lambda_{2,3}$ are negative if $\frac{\Lambda\sigma}{m+\sigma}>K\mu \Rightarrow R_{h}>1.$ This shows that $E_{f}$ is locally asymptotically stable if $R_{h}>1$ and $R_{m}<1.$ In this case, bee keepers should carry out practices that would bring $R_{m}$ below unity in addition to practices that lead to $R_{h}>1.$ This requires that the bee-keepers carry out practices that aim at reducing the parasitism rate, $\alpha$ and conversion rate of nutrients, $c$ to support mite reproduction. Bee-keeping practices that aim at increasing the total mite mortality rate, $\mu_{3}$ and those that would reduce the queen,s egg-laying rate, $\Lambda$ are also recommended to this effect. Such practices include those intended to directly kill the mites and those that limit mite reproduction. The mite control methods that directly kill the mites include; sprinkling powdered sugar on the adult bees to induce grooming of mites, mite trapping intended to kill reproductive mites, using screened bottom boards to kill mites that fall on the bottom board and spraying the bees with organic varroacides to kill both phoretic and reproductive mites. Those that limit mite reproduction include; caging the queen bee to reduce on the brood cells that facilitate mite reproduction, using a young queen to reduce on drone brood thus limiting mite reproduction and keeping Varroa Sensitive Hygiene (VSH) bees to kill reproductive mites inside the pupae cells which also limits mite reproduction.
{\thm The mite-infested equilibrium ($E_{i}$) of Model (1) is locally asymptotically stable when $R_{m}>1.$}\\
{\bf Proof:} The proof is supported by the following Lemma (MacCluskey and van den Driesche, 2004).
{\lem Let $M$ be a $3\times3$ matrix. If trace $(M),$ determinant $(M)$ and determinant $(M^{[2]})$ are all negative then all eigenvalues of $M$ have negative real parts.}\\
Now, From the Jacobian (J) of System $(1)$ given by Equation $(8),$ we have;
$$ trc\, (J_{E_{i}})=-(\frac{\Lambda\sigma c\alpha}{Kc\alpha\mu+\sigma\mu_{3}}+\mu)<0~~~~\mbox{and}~~~ det\,(J_{E_{e}})=-c\alpha^{2}\mu x^{*}z^{*}<0$$
The second additive compound matrix $J_{E_{i}}^{[2]}$ of System $1$ using the Jacobian is given as
 \begin{eqnarray}
 J_{E_{i}}^{[2]}=\left[\begin{array}{ccccc}-(m+\sigma+\alpha z^{*}+\mu)&0&\alpha x^{*}\\ 0 &-(m+\sigma+\alpha z^{*}) & \frac{\Lambda K}{(K+y^{*})^{2}} \\ -c\alpha z^{*}&\sigma&-\mu \end{array}\right]
\end{eqnarray}
from which $$ det(J_{E_{i}}^{[2]})=-(m+\sigma+\alpha z^{*})[(m+\sigma+\alpha z^{*}+\mu)\mu+c\alpha^{2} z^{*}] <0$$ implying that System $(1)$ has a local stability at $E_{i}.$
\subsubsection{Global Stability Analysis of the Mite-Free Equilibrium $E_{f}$}
{\thm The mite-free equilibrium ($E_{f}$) of Model (1) is globally asymptotically stable when $R_{m}<1$.}\\
{\bf Proof:} X=(x,y) and Y=(z) that represent the mite-free and mite-infested compartments, respectively are obtained following the recommended form by Castillo-Chavez\emph{ et al}., (2002). Therefore,

\begin{eqnarray*}
 F(X,0)=\left[\begin{array}{ccccc}\frac{\Lambda y}{K+y}-(m+\sigma)x\\ \sigma {x}-\mu{y} \\ 0 \end{array}\right] ~~\mbox{and}~~~G(X,Y)=\left[\begin{array}{ccccc}c\alpha xz-\mu_{3}z\end{array}\right],~~\mbox{with}~~~G(X,0)=0
\end{eqnarray*}
The Jacobian of
$G(X,Y)|_{E_{f}}\,\,\mbox{is}\,\,
A=  \left(
   \begin{array}{c}
     c\alpha x^{*}-\mu_{3}\nonumber
   \end{array}
 \right).$
%\end{equation}
Then $G(X,Y)$ can be written as  $G(X,Y)=AY-\hat {G}(X,Y)$ where
$\hat {G}(X,Y)=c\alpha z(x^{*}-x)>0$ since $x^{*}=x|_{E_{f}}>x.$ This implies that the condition H2 of the theorem by Castillo-Chavez \emph{et al}., (2002) is satisfied. Also, as $t\rightarrow\infty,$  $X^{*}\rightarrow E_{f}(\frac{\Lambda}{m+\sigma}-\frac{K\mu}{\sigma},\frac{\Lambda\sigma}{\mu(m+\sigma)}-K,0)$
regardless of the initial value $x(0)$ and $y(0)$. In order to exclude the limit cycles, we multiply each equation of $X$ by a Dulac multiplier $\frac{1}{xy}$ so that;
$$\frac{\partial}{\partial x}(\frac{\Lambda}{x(K+y)}-\frac{\sigma}{y})+\frac{\partial}{\partial y}(\frac{\sigma}{x}-\frac{\mu}{x})=-(\frac{\Lambda}{x^{2}(K+y)}+\frac{\sigma}{y^{2}})<0$$
which rules out periodic orbits. Therefore, from the Poincare-Bendixson Theorem, $(X^{*},0)$ is globally asymptotically stable when $R_{m}<1.$  This means that it is possible to eradicate mites from a bee colony if $R_{m}<1$. Therefore, bee-keepers should carry out appropriate practices so that $R_{m}$ goes below unity in order to eradicate mites from a colony. Such practices would target reducing parasitism rate, $\alpha$ and conversion rate of nutrients to support mite reproduction, $c,$ increasing the total mite mortality rate, $\mu_{3}$ and those that would reduce the queen's egg-laying rate $\Lambda.$ Bee keepers may also remove drone brood (pupae) from the colony since mites prefer to reproduce in drone cells than in worker cells. This will limit mite reproduction opportunities thus reducing mite population growth. In addition, increasing $\mu_{3}$ requires that a bee keeper selects a good breed of bees that are highly hygienic and can effectively groom mites off their bodies.
\subsubsection{Global Stability Analysis of the Mite-Infested Equilibrium $E_{i}$}
\paragraph\ We now establish conditions under which in the long run both the mites and honey bees settle at an equilibrium and how this is independent of the initial number of mites invading the colony. Given the Jacobian (J), of System (1) and a diagonal matrix (D) with entries $\pm1$ then, it can be shown that the matrix DJD has both positive and negative off-diagonal elements implying that System $(1)$ is not competitive or monotone. This implies that the method of monotone dynamical systems cannot be applied to study the global stability of $E_{i}$ (Li and Muldowney, 1995). This is because such systems do not obey the Poincare-Bendixson property. In this case, Liao and Wang (2012) suggest a Volterra-Lyapunov stable matrix method to study the global stability of $E_{i}.$\\\\
\textbf{The Volterra-Lyapunov Global Stability Analysis of  $E_{i}$}
\paragraph\ We use the classical Lyapunov method combined with Volterra-Lyapunov stable matrices to study the global stability of $E_{i}$ (Tian and Wang, 2011; Liao and Wang, 2012; Chien and Shateyi, 2021). The following definitions/propositions for the volterra-lyapunov stability criterion as given in Chien and Shateyi (2021) are highlighted below:\
{\defn
\begin{itemize}  
 \item[(i)] {\it If there exists a positive diagonal matrix $D_{n\times n}>0$ such that $DA+(DA)^{T}<0,$  then $A_{n\times n}$ is Volterra-Lyapunov stable.} 
 \item[(ii)] {\it If there exists a diagonal matrix $D_{n\times n}>0$ such that $DA+(DA)^{T}>0,$  then $A_{n\times n}$ is diagonally stable.}
\end{itemize}}
{\prop   {The matrix $D_{2\times 2}=[d_{ij}]:i,j=1,2$ is Volterra-Lyapunov stable if and only if $d_{11}<0,d_{22}<0$ and $det(D)=d_{11}d_{22}-d_{21}d_{12}>0.$}}\\
{\prop   {Let $D_{n\times n}=[d_{ij}]$ for $n\geq2$, the positive diagonal matrix $E_{n\times n}=diag(e_{1},e_{1},...,e_{n})$ and $F=D^{-1},$ such that: $d_{nn}>0,$ $\tilde{E}\tilde{D}+(\tilde{E}\tilde{D})^{T}>0$ and $\tilde{E}\tilde{F}+(\tilde{E}\tilde{F})^{T}>0$, then there is $e_{n}>0$ such that $ED+(ED)^{T}>0.$ Note that we denote the matrix resulting from deleting the last row and last column of the matrix $D$ by $\tilde{D}_{(n-1)\times (n-1))}.$}}\\ 
\paragraph\ System $(1)$ is studied in the biologically feasible domain $\omega$=\{(x,y,z)$\in R^{3}_{+}:0\leq c{(x+y)}+{z}\leq \frac{c\Lambda}{\gamma}\}$ which is positively invariant in $R^{3}_{+}$ as highlighted in sub-subsection 2.2.1. We consider the following Lyapunov function;
$$V=w_{1}(x-x^{*})^{2}+w_{2}(y-y^{*})^{2}+w_{3}(z-z^{*})^{2}$$
where $w_{1},w_{2}$ and $w_{3}$ are positive constants. The derivative of $V$ along the solutions of System $(1)$ is
\begin{eqnarray*}
\frac{dV}{dt} &=& 2w_{1}(x-x^{*})\dot x+2w_{2}(y-y^{*})\dot y+2w_{3}(z-z^{*})\dot z\\
&=& 2w_{1}(x-x^{*})[\frac{\Lambda y}{K+y}-\frac{\Lambda y^{*}}{K+y^{*}}-(\sigma+m)(x-x^{*})-\alpha xz+\alpha x^{*}z^{*}]\\
&+& 2w_{2}(y-y^{*})[\sigma(x-x^{*})-\mu(y-y^{*})]+2w_{3}(z-z^{*})[c\alpha xz-\mu_{3}(z-z^{*})-c\alpha x^{*}z^{*}].
\end{eqnarray*}
It is clear that when $x=x^{*}, y=y^{*}$ and $z=z^{*}$ then $\frac{dV}{dt}=0.$ We seek to show that when $x\neq x^{*},y\neq y^{*}$ and $z\neq z^{*},$ then $\frac{dV}{dt}<0.$ Adding and subtracting expressions $\alpha xz^{*}$ and $c\alpha xz^{*}$ in the first and third square brackets, respectively yields;
\begin{eqnarray*}
\frac{dV}{dt} &=& 2w_{1}(x-x^{*})[\frac{\Lambda K(y-y^{*}) }{(K+y)(K+y^{*})}-(\sigma+m)(x-x^{*})-\alpha z^{*}(x-x^{*})-\alpha x(z-z^{*})]\\
&+& 2w_{2}(y-y^{*})[\sigma(x-x^{*})-\mu(y-y^{*})]+ 2w_{3}(z-z^{*})[c\alpha z(x-x^{*})-\mu_{3}(z-z^{*})+c\alpha x^{*}(z-z^{*})].
\end{eqnarray*}
On simplification, we have
\begin{eqnarray}
\frac{dV}{dt}&=&2w_{1}[\frac{\Lambda K }{(K+y)(K+y^{*})}(y-y^{*})(x-x^{*})-(\sigma+m+\alpha z^{*})(x-x^{*})^{2}-\alpha x(z-z^{*})(x-x^{*})]\nonumber\\
&+&2w_{2}[\sigma(x-x^{*})(y-y^{*})-\mu(y-y^{*})^{2}]\\
&+& 2w_{3}[c\alpha x(z-z^{*})^{2}+c\alpha z^{*}(x-x^{*})(z-z^{*})-\mu_{3}(z-z^{*})^{2}].\nonumber
\end{eqnarray}
Expressing the lyapunov derivative in Equation (10) in matrix form gives
$$\frac{dV}{dt}=L(MQ+Q^{T}M^{T})L^{T}$$
where $L=[x-x^{*},y-y^{*},z-z^{*}],$ $M=\mbox{diag}\{w_{1},w_{2},w_{3}\}$ and
\begin{eqnarray}
 Q=\left[\begin{array}{ccccc}-(m+\sigma+\alpha z^{*})&\frac{\Lambda K}{(K+y^{*})(K+y)}& -\alpha{x}\\ \sigma & -\mu & 0 \\ c\alpha z^{*} &0&c\alpha x-\mu_{3} \end{array}\right].
\end{eqnarray}
To show that $E_{i}$ is globally asymptotically stable, we need to prove that $Q$ is Volterra-Lyapunov stable. The following steps that follow from Proposition $2.2.2$ and  outlined in Chien and Shateyi (2021) are used to prove that $Q$ is Volterra-Lyapunov stable: (i) $-Q_{33}>0$ (ii) $\widetilde{-Q}$ is diagonally stable and (iii) $\widetilde{-Q^{-1}}$ is diagonally stable.
{\thm Suppose Equation $(11)$ defines matrix $Q_{3\times 3},$ then $Q_{3\times 3}$ is Volterra-Lyapunov stable.}\\
{\bf Proof:} It is clear that $-Q_{33}>0$ if $\frac{\mu_{3}}{c\alpha}> x.$ The left hand side of this inequality can be written in terms of $R_{m}$ and $R_{h}$ using expressions $(6)$ and $(7)$ to give
\begin{equation}
 \frac{K\mu(R_{h}-1)}{R_{m}\sigma}>x
\end{equation}
The right hand side of inequality (12) is positive since it represents the pupae population. Therefore, inequality $(12)$ holds if its left hand side is positive which leads to the conditions;
\begin{equation}
R_{h}>1\,\,\mbox{and}\,\,R_{m}>1.
\end{equation}
Inequality $(12)$ would also mathematically hold if $R_{m}>0.$ However, this would include the case, $0<R_{m}<1$ for mite extinction which contradicts the existence of $E_{i}$ and therefore discarded. We proceed to show that $\widetilde{-Q}$ is diagonally stable. Deleting the last row and last column of $-Q$ gives $\widetilde{-Q}.$ It follows that;
\begin{eqnarray*}
 \widetilde{-Q} =\left[\begin{array}{ccccc}(m+\sigma+\alpha z^{*})&-\frac{K\Lambda}{(K+y)(K+y^{*})}\\ -\sigma & \mu \end{array}\right]
\end{eqnarray*}
It is clear that $\widetilde{-Q_{11}}>0,$ $\widetilde{-Q_{22}}>0$ and
\begin{equation}
det(\widetilde{-Q})=(m+\sigma+\alpha z^{*})\mu-\frac{K\Lambda\sigma}{(K+y)(K+y^{*})}.
\end{equation}
The first two equations of Model $(1)$ evaluated at $E_{i}$ are solved simultaneously to get
\begin{equation}
(m+\sigma+\alpha z^{*})\mu=\frac{\Lambda\sigma}{K+y^{*}}.
\end{equation}
Simplifying the determinant by substituting Eqn $(15)$ into Eqn $(14)$ and then factoring out $\frac{\Lambda\sigma}{K+y^{*}}$ yields;
\begin{eqnarray*}
 det(\widetilde{-Q})=\frac{\Lambda\sigma y}{(K+y)(K+y^{*})}>0.
\end{eqnarray*}
Therefore, $\widetilde{-Q}$ is diagonally stable. 
\paragraph\ Lastly, it is shown that $\widetilde{-Q^{-1}}$ is diagonally stable.
Defining $\widetilde{-Q^{-1}}$ by deleting the last row and last column of $Q^{-1}$ gives;
\begin{eqnarray*}
 \widetilde{-Q^{-1}} = \frac{1}{-det(Q)} \left[\begin{array}{ccccc}\mu(\mu_{3}-c\alpha x)&\frac{K\Lambda}{(K+y)(K+y^{*})}(\mu_{3}-c\alpha{x})\\ \sigma(\mu_{3}-c\alpha {x}) & (m+\sigma_{1}+\alpha z^{*})(\mu_{3}-c\alpha x)+c\alpha^{2} xz^{*} \end{array}\right].
\end{eqnarray*}
In addition, we obtain
\begin{eqnarray*}
-det{(Q)}= (\mu_{3}-c\alpha x)[\mu(m+\sigma_{1}+\alpha z^{*})+\frac{\Lambda\sigma}{(K+y)(K+y^{*})}]+c\alpha^{2}\mu xz^{*}>0
\end{eqnarray*}
and hence, it follows that
\begin{eqnarray*}
det(\widetilde{-Q^{-1}})&=&\frac{1}{-det{(Q)}}[(\mu_{3}-c\alpha x)^{2}\{(m+\sigma+\alpha z^{*})\mu-\frac{K\Lambda\sigma}{(K+y)(K+y^{*})}\}+(\mu_{3}-c\alpha x)c\alpha^{2}\mu xz^{*}]\\
&=&\frac{1}{-det{(Q)}}[(\mu_{3}-c\alpha x)^{2}\frac{\Lambda\sigma y}{(K+y)(K+y^{*})}+(\mu_{3}-c\alpha x)c\alpha^{2}\mu xz^{*}]>0.
\end{eqnarray*}
Therefore, $\widetilde{-Q^{-1}}$ is diagonally stable. This implies that $Q$ is Volterra-Lyapunov stable. We now provide the following result/theorem:
{\thm The mite-infested equilibrium $E_{i}$ of Model $(1)$ is globally stable when $R_{h}>1$ and $R_{m}>1$.}\\
{\bf Proof:} Theorem $2.2.4$ concludes that there exists a positive diagonal matrix $M$ such that\\  $MQ+Q^{T}M^{T}<0$. Therefore, $\frac{dV}{dt}<0$  when $x\neq x^{*},y\neq y^{*}$ and $z\neq z^{*},$ which ensures the global stability of $E_{i}.$
\paragraph\ The global stability analysis presented above shows that without proper interventions by the bee-keeper, the mites persist in the colony forever. In this case, a bee-keeper should carry out practices that reduce the mite reproduction number, $R_{m}$ to a value below unity while at the same time carry out practices that increase the basic demographic reproduction number, $R_{h}$ to a value above unity. Reducing the egg-laying rate $(\Lambda)$ reduces both $R_{m}$ and $R_{h}$. The egg-laying rate $(\Lambda)$ can be reduced by caging the queen bee which reduces the available brood cells for mite reproduction. To ensure colony survival, caging the queen bee should be done for a short period of time so that $R_{h}>1$ which ensures bee population growth.
\paragraph\ In addition, a bee keeper can focus on reducing the brood cells so as to limit mite reproduction opportunities. Whereas, reducing the brood cells reduces mite reproduction opportunities, it may deplete the colony of the brood cells. This leads to $R_{h}<1$ which causes colony extinction. Therefore, bee-keepers are advised to remove drone brood only because they harbour more mites than worker brood. In addition, removing drone brood does not affect colony performance since drone bees do not participate in colony work.
\paragraph\ Equation $(7)$ shows that increasing the half saturation constant $K$ reduces $R_{m}.$ This slows down mite population growth thus, improving colony performance. This is because bigger colonies (which imply big values of $K$), produce less brood per bee which limits mite-reproduction opportunities (Harbo, 1986). In this case, a bee-keeper is advised to increase the strength of weak colonies which increases the value of $K$. This can be done by transferring combs with bees from strong colonies to weak colonies or uniting at least two weak colonies. This increases colony strength and hence the value $K$ which reduces $R_{m}$ thus, increasing colony performance. It should however, be noted that the increase in colony strength which implies an increase in the value of $K,$ results in $R_{h}<1$ (Equation $(6)$) thus reducing bee population growth. This is because a big number of adult bees that care for the queen $(K)$ consume large amounts of honey and large quantities of pollen resulting into a reduction in the queen's food. This negatively affects the queen's capacity to lay eggs which reduces colony performance. A bee keeper must therefore decide whether to increase $K$ and reduce $R_{m}$ to a value below unity i.e., $R_{m}<1$ (wiping out the mites from the colony) or increase $K$ and reduce $R_{h}$ to a value below unity i.e., $R_{h}<1$ (reducing bee population growth). This leads to the so-called ``paradox of enrichment." To overcome this paradox, bee-keepers must maintain an optimal number of adult bees $(K)$ in the colony that care for the queen bee, to ensure colony survival $(R_{h}>1)$ and at the same time reduce mite population growth $(R_{m}<1)$.
\paragraph\ Increasing the value of the total mite mortality rate $(\mu_{3})$ reduces $R_{m}$ hence reducing mite population growth. In this case, bee-keepers may opt to keep adult bees that are highly hygienic and can effectively groom mites off their bodies.
Reducing the rate at which the reproductive mites invade the pupae cells $\alpha,$ reduces $R_{m}$ hence reducing mite population growth. Therefore, a bee-keeper is advised to apply varroacides in order to kill the mites. However, this strategy must be used sparingly to avoid contaminating honey with the chemicals that are poisonous to the bees and human life. In this case, bee-keepers should apply soft chemicals such as formic acid. This is because formic acid has sufficient efficacy against the mites and can kill the mites within the brood cells (Fries \emph{et al}., 1991). Besides, formic acid having a low probability of eliciting resistance after being used, it also has a low risk of residues which reduces the chance of honey contamination. Similarly, an increase in the conversion rate of nutrients to support mite reproduction, $(c)$ increases $R_{m}$ thus, increasing mite population growth. Therefore, any practice by the bee-keeper that would reduce $c$ is recommended.
\paragraph\ From Equation $(7),$ it follows that  $\lim\limits_{\sigma\rightarrow\infty}R_{m}\rightarrow 0$ and $\lim\limits_{\sigma\rightarrow0}R_{m}\rightarrow \infty.$ This means that increasing the emergence rate of adult bees from the pupae $(\sigma)$, reduces $R_{m}$ whereas reducing it increases $R_{m}.$  Increasing $\sigma$ shortens the pupae hatching period. This reduces the chances for mites to mature inside the pupae cells thus, reducing mite population growth ($R_{m}$). Therefore, bee-keepers should choose interventions that increase the adult bees that care for the queen $(K).$ These interventions aim at increasing the strength of colonies which can be achieved by uniting at least two weak colonies or transferring bees from strong colonies to weak colonies in order to increase their strength. This is because increasing the adult bees that care for the brood and queen $(K)$ speeds up the maturation rate of the pupae. This leads to the birth of immature daughter mites which cannot survive outside the pupae cells ultimately increasing mite mortality. The process of uniting colonies could lead to the transfer of mites from one colony to another and must be done carefully. Increasing the adult bee mortality rate $(\mu)$ would imply that the adult bees live for a short time. This is not desirable and therefore not recommended. Once this is done it leads to colony collapse.
\subsection{Evaluation of Varroa Mite Control Methods}
\paragraph\ In this section, an evaluation of varroa control strategies is carried out. This is done through a sensitivity analysis on the model parameters. This helps in advising bee keepers on the suitable varroa control methods so as to increase honey bee colony performance.
\subsubsection{Sensitivity Analysis}
\paragraph\ Sensitivity analysis is carried out to identify the parameters that have a high effect on the basic mite reproduction number $(R_{m})$. This enables us to design effective mite control strategies. We use the Normalized Sensitivity index (NSI) to identify these parameters. The NSI measures the relative change of $R_{m}$ with respect to a parameter for example, $\omega$ and is given by;
$$ \Upsilon^{R_{m}}_{\omega}=\frac{\partial R_{m}}{\partial \omega}\times \frac{\omega}{R_{m}}$$
\begin{table}[h]
\vspace{1.3cm}
\centering
\caption{ \it The computed Normalised Sensitivity Indices for $R_{m}$} \label{tab1a}
\vspace{0.6cm}
\begin{tabular}{lllll}
\toprule
{ Parameter/Symbol}&{ Parameter value}&  	{  Index for $R_{m}$}\\ 
\midrule
~~~~~~~~~~~~~$\Lambda$&~~~~~~~~~~1500&~~~+1.1559889   \\ 
~~~~~~~~~~~~~$\alpha$&~~~~~~~~~~[0,1]&~~~+1.0000000	  \\
~~~~~~~~~~~~~$c$&~~~~~~~~~~[0,1]&~~~+1.0000000          \\
~~~~~~~~~~~~~$d_{m}$&~~~~~~~~~~0.006&~~~-0.0441176     \\
~~~~~~~~~~~~~$K$&~~~~~~~~~~5000&~~~-0.1559889  \\

~~~~~~~~~~~~~$\mu$&~~~~~~~~~~0.04&~~~-0.1559889    \\
~~~~~~~~~~~~~$g$&~~~~~~~~~~0.05&~~~-0.3676471   \\
~~~~~~~~~~~~~$h$&~~~~~~~~~~0.08&~~~-0.5882353    \\
~~~~~~~~~~~~~$\sigma$&~~~~~~~~~~0.083&~~~-0.9862382\\
\bottomrule
\end{tabular}
\end{table}
\paragraph\ We first computed the Normalised Sensitivity Indices to determine the effect of a change of $10\%$ in the parameters on the mite basic reproduction number, $R_{m}.$ It can be observed from Table $2$ that increasing (decreasing) the egg laying rate $(\Lambda)$ and the parasitism rate $(\alpha)$ by $10\%$ while keeping the rest of the parameters constant, increases (decreases) the values of $R_{m}$ by $11.6\%$ and $10\%$ respectively. On the other hand, a $10\%$ increase (decrease) in the emergence rate of adult bees from the pupae $\sigma$ while keeping the rest of the parameters constant, decreases (increases) the mite reproduction number, $R_{m}$ by $9.9\%.$ It is deduced that honey bee mite-infestation increases when parameter values of $\Lambda,$ $\alpha$ and $c$ are increased and/or those of $h,$ $g,K,\sigma,\mu$ and $d_{m}$ are decreased. 
\paragraph\ The computed Normalised Sensitivity Indices as given in Table 2 show that parameter $\Lambda$ has the highest effect on $R_{m}$ followed by $\alpha$ and $c.$ This means that the egg-laying rate is the most significant determinant of a mite outbreak in a colony. Equivalently, this means that the number of brood cells in the colony determines mite population growth. This implies that control methods that focus on reducing the egg-laying rate, $\Lambda$ or reducing the brood cells in the colony quickly reduce $R_{m}$ to a value below unity.  This is because a reduction in $\Lambda$ reduces the number of brood cells in the colony which limits mite reproduction opportunities. The mite control methods aimed at limiting mite reproduction by reducing $\Lambda$ or reducing the brood cells include; caging the queen bee, using a young queen bee and trapping mites using artificial drone combs. Therefore, these three mite control strategies are considered to be the most effective strategies for reducing $R_{m}$ to a value below one.
\paragraph\ Caging the queen temporarily stops the queen from laying eggs. This reduces the brood cells thus, denying the mites the chance to reproduce. On the other hand, young queens produce less drone brood which limits mite reproduction. This is because mites have a distinct preference to reproduce in drone brood cells than in worker brood cells. Therefore, less drone brood implies less mite-reproduction opportunities which reduces mite population growth.
\paragraph\ Bee keepers can also trap mites using artificial drone-combs. These drone combs are then removed from the colony and the brood scrapped off the combs or freezed to kill the mites. This reduces the drone brood cells and kills reproductive mites inside the drone pupae cells thus, limiting mite reproduction opportunities.
In addition, controls that target mite mortality, $d_{m}$ emergence rate of adult bees from pupae cells, $\sigma$ hygienic rate, $h$ and grooming rate, $g$ also reduce the mites in a colony.
\paragraph\ Currently, bee-farmers use various interventions that target certain parameters aimed at reducing mite population growth. For example, they introduce young queens to increase the number of eggs laid by the queen but with less drone brood. Other control interventions practiced by bee-farmers include: caging the queen bee to cause a brood break so as to reduce mite reproduction, sprinkling powdered sugar onto the bees to encourage grooming, using screened bottom boards to reduce mite invasion into brood cells, mite-trapping using drone combs to limit mite reproduction, spraying the bees with organic chemicals to kill the mites and keeping Varroa Sensitve Hygiene(VSH) bees to remove dead and mite-infested brood. These intervention strategies affect different parameters. We determine the parameters that are affected by the different interventions so that an evaluation of their effectiveness is documented. For example caging the queen temporarily for about three weeks affects the egg laying rate, $\Lambda$ which subsequently influences the parasitism rate, $\alpha.$ A temporary brood break also increases the phoretic mite population which encourages grooming, $g.$ Similarly, introducing young queens affects $\Lambda,K$ and $\alpha;$ mite-trapping using drone brood combs affects $\alpha;$ sprinkling bees with powdered sugar affects $g$ and $\alpha;$ adding screened bottom board affects $\alpha;$ spraying bees with organic chemicals affects $\alpha$ and keeping VSH bees affects $h,\alpha.$
{\defn ({\emph {Bakare}} {et al.,} {\emph {2020}}) {Total efficacy of an intervention strategy is the overall sum of the protection provided by each intervention strategy that is employed to eliminate a disease or an outbreak.}}
\paragraph\ For example, introducing a young queen bee produces more brood with less drone brood which affects the egg laying rate, $\Lambda.$ They also produce stronger colony populations which influences the number of adult bees, $K$ that care for the queen bee. Such colonies that have young queens have low mite infestation levels thus, reducing parasitism $\alpha$ (Aky$\ddot{o}$l \emph{et al}., 2007). Therefore, the total efficacy is the absolute sum of all the elasticities corresponding to the affected parameters and is computed as; $1.1559889+1-0.1559889=2.$ Similarly computation of the total efficacy of the different mite-control methods is done and  summarised as follows (Table $3$).
\begin{table}[h!]
\caption{Intervention strategies and their efficacy}
\centering
\begin{tabular}{llll}
  \toprule
  Intervention& affected parameter(s)&Total efficacy&Rank\\
  \midrule
Introducing a young queen bee & $\Lambda,\alpha,K$&2.0000000&1\\
Caging the queen bee & $\Lambda,\alpha,g$&1.7883418&2\\
Mite-trapping & $\alpha$ &1.0000000&4\\
Adding screened-bottom boards & $\alpha$ &1.0000000&4\\
Spraying the bees with organic chemicals &$\alpha$ &1.0000000&4\\
Sprinkling the bees with powdered sugar &$g,\alpha$ &0.6323529&6\\
Keeping VSH Bees &$h,\alpha$ &0.4117647& 7\\
\bottomrule
\end{tabular}
\end{table}
\paragraph\ The total efficacy computations in Table $3$ show that introducing a young queen bee is the most effective strategy followed by caging the queen bee. It should be noted that the first two control strategies are those that target the egg-laying rate, $\Lambda$ of the queen bee. This is in agreement with the sensitivity analysis results indicating that $\Lambda$ has the highest effect on reducing the basic mite reproduction number,  $R_{m}.$ Because caging the queen bee is sometimes risky, and bee-keepers may fear trying it out, mite-trapping, adding screened-bottom boards, spraying the bees with organic chemicals are other possible recommended effective control strategies (Table $3$). Sprinkling the bees with powdered sugar and keeping Varroa Sensitive Hygienic bees should be done as a last resort due to their low total efficacy as shown in Table $3.$
\section{Numerical Simulations of the Model}
\subsection{Long Term Behavior of the Model}
\paragraph\ In this section, numerical simulations using parameter values in Table $1$ are carried out to confirm the analytical results. We simulated the model over a period of 500 days since the life span of the queen bee lies between $1-3$ years. Initial adult bee and pupae populations used in the simulations are chosen such that the ratio of adult bees to pupae cells is approximately 2 as in Torres and Torres (2020). An illustration of the global stability of the mite-free equilibrium (Fig.2(a)) and mite-infested equilibrium (Fig.2(b)) is presented.
\begin{figure}[H]
 \begin{minipage}{0.3\textwidth}
  \centering
  \includegraphics[width=9cm]{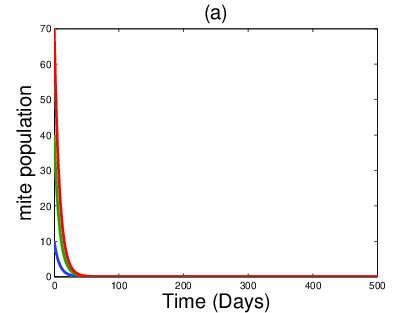}
 \end{minipage}\hfil
\begin{minipage}{0.3\textwidth}
  \centering
  \includegraphics[width=9cm]{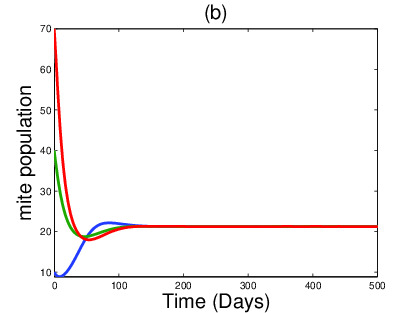}
 \end{minipage}\hfil
\caption{\it Time series plots for $x(0)=7000,y(0)=15000$ and $10\leq z\leq70$ for both figures. Parameters used are found in Table $1$ with $c=\alpha=0.002$ (so that $R_{m}=0.46<1$) for Fig.(a) and $c=\alpha=0.007$ (so that $R_{m}=5.6>1$) for Fig.(b).}
\end{figure}

\paragraph\ Results of Figure $2(a)$ show that it is possible to eliminate the mites from a mite-infested colony when $R_{m}<1$ (Theorem $2.2.3$). Results of Figure $2(b)$ are also in agreement with Theorem $2.2.5$ which states that it is possible for the mites to co-exist with the bees regardless of the number of mites infesting the colony if $R_{m}>1.$ In this case all trajectories converge towards the endemic equilibrium. Therefore, bee-keepers need to apply control methods that target parameters in $R_{m}$ (Equation $(7)$) so that $R_{m}$ goes below unity for mite elimination. Some of the methods include: caging the queen bee in order to target the egg-laying rate $(\Lambda)$ and spraying the bees with organic chemicals to target the parasitism rate $\alpha.$
\subsection{Effect of Mite-Control Strategies}
\paragraph\ In this section, numerical simulations are carried out to investigate the effect of the commonly used mite-control strategies on honey bee colony performance.
\subsubsection{Caging the Queen Bee/Using a Young Queen Bee}
\paragraph\ These control strategies are used by bee-keepers who intend to limit mite-production. Caging the queen temporarily stops the queen from laying eggs. Therefore, this mite-control strategy directly affects the egg-laying rate parameter $\Lambda$ in Model $(1).$ Also, bee-keepers who use young queens intend to increase $\Lambda$ so as to produce more worker brood, less drone brood and produce stronger colony populations (Aky$\ddot{o}$l \emph{et al}., 2007). We simulate the effect of these two mite-control strategies on the population dynamics of bees and mites by varying $\Lambda$ in Model $(1)$ in the range $600-1000.$ The parameters used are found in Table $1$ with $c=0.0075,\alpha=0.0073.$
\begin{figure}[H]
\begin{minipage}{0.3\textwidth}
  \centering
  \includegraphics[width=6cm]{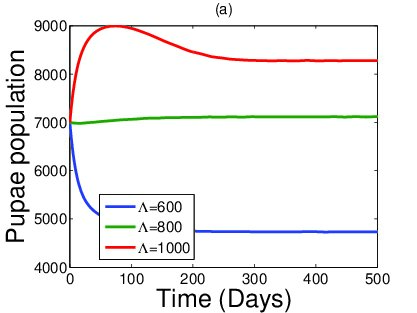}
\end{minipage}\hfil
\begin{minipage}{0.3\textwidth}
  \centering
  \includegraphics[width=6cm]{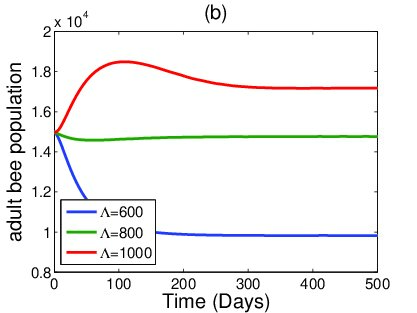}
\end{minipage}\hfil
\begin{minipage}{0.3\textwidth}
  \centering
  \includegraphics[width=6cm]{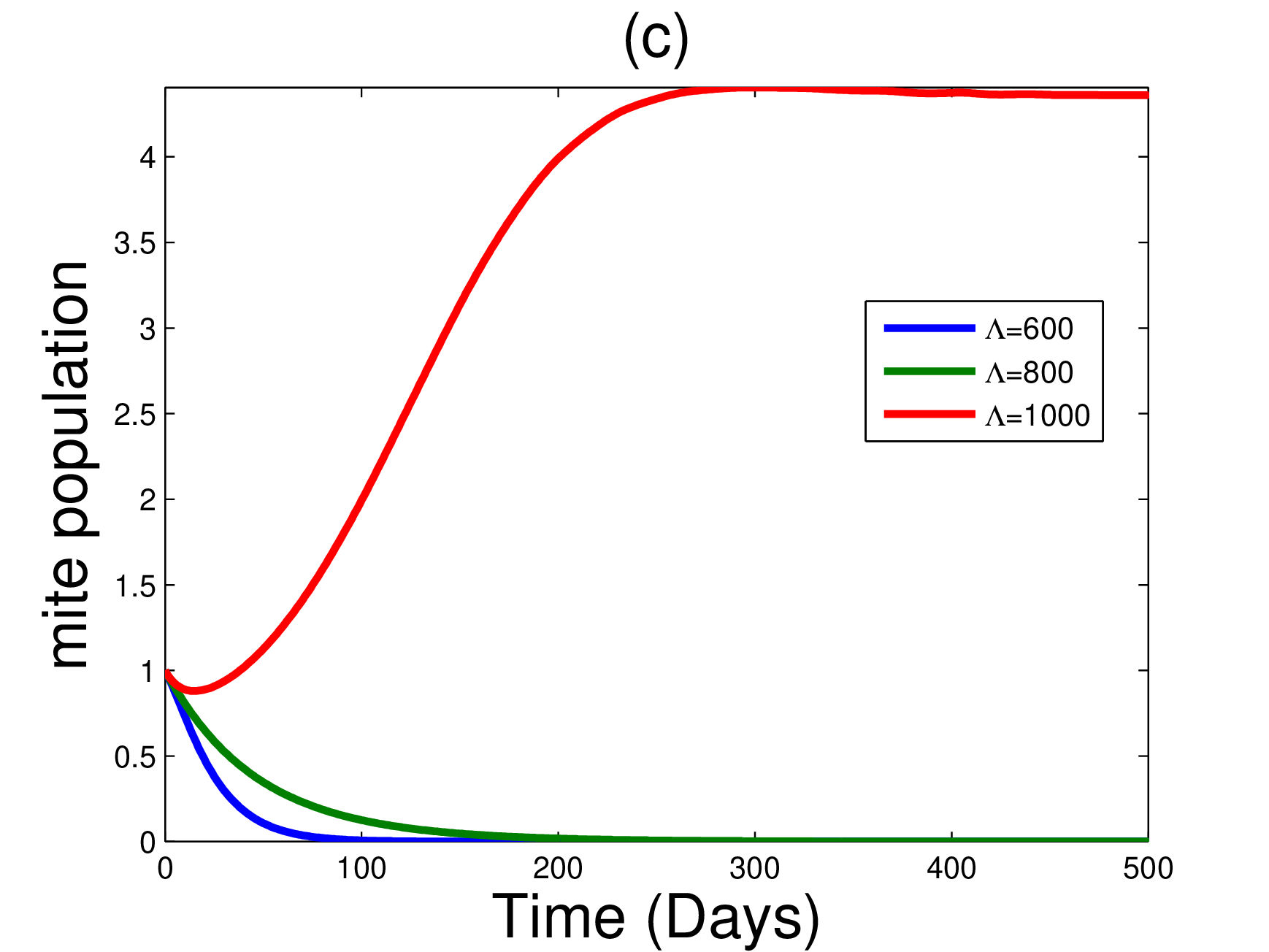}
\end{minipage}\hfil
\caption{\it Time-varying Plots of pupae cells, adult bees and mites with variations of the egg-laying rate of the queen bee with initial conditions: $x(0)=7000,y(0)=15000,z(0)=1.$}
\end{figure}
\paragraph\ Results of Figure $3$ show that the pupae, adult bee and mite populations decrease with decreasing values of $\Lambda$ (Equation $(1)$ of Model $(1)$).  From the first equation of Model $(1),$ a decrease in $\Lambda$ reduces the number of pupae cells $\frac{\Lambda y}{K+y}$ which limits mite reproduction opportunities thus reducing the parasitism rate $(\alpha).$ The reduction in $\alpha$ increases the average life span of the pupae cells which increases their number. This consequently increases the adult bees thus increasing colony performance. Therefore, by consistently applying any of the two control methods, the mite population can be eliminated and colony performance increased. However, caging the queen must be done for a short period to avoid exhausting the brood cells which causes colony extinction. In this case, using a young queen may be an appropriate method that provides long term or sustainable solutions to the mite-infestation problem.
\subsubsection{Sprinkling Adult Bees with Powdered Sugar}
\paragraph\ Sprinkling adult bees with powdered sugar induces the adult bees to groom mites off their bodies. This control strategy affects the grooming rate $(g)$ of the adult workers in Model $(1).$ Torres and Torres (2020) observed that adult workers have the capability of grooming off their bodies $5\%$ of the mites per day (Table $1$). Therefore, the effect of grooming on the colony is investigated by varying $g$ in the range $1\%-9\%$ per day. Results from Figures $4(a)$ and $4(b)$ show that increasing $g$ increases both the pupae and adult bee population. On the other hand, the mite population reduces with increasing values of $g.$ From the third equation of Model $(1),$ an increase in $g$ decreases the average life span, $\frac{1}{h+g+d_{m}}$ of the mites. This accelerates mite mortality thus, reducing parasitism rate $(\alpha)$. The reduction in $\alpha$ increases the average life span, $\frac{1}{\sigma+m+\alpha z}$ of the pupae cells (Equation $(1)$ of Model $(1)$). This consequently increases the adult bees thus, increasing colony performance. Therefore, in addition to selecting a breed of bees that can effectively groom mites off their bodies, bee-keepers may induce bee grooming by sprinkling adult bees with powdered sugar.
\begin{figure}[H]
\begin{minipage}{0.3\textwidth}
  \centering
  \includegraphics[width=6cm]{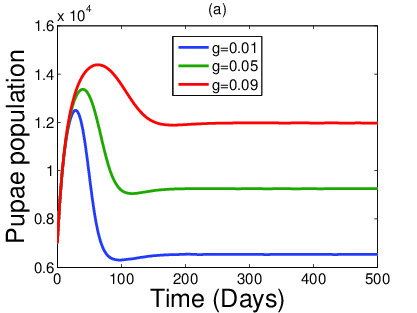}
\end{minipage}\hfil
\begin{minipage}{0.3\textwidth}
  \centering
  \includegraphics[width=6cm]{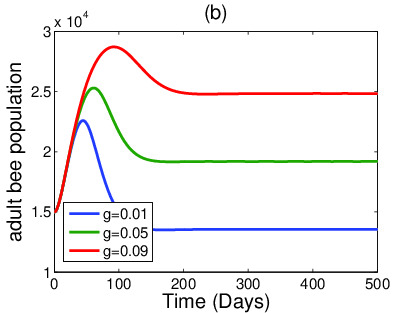}
\end{minipage}\hfil
\begin{minipage}{0.3\textwidth}
  \centering
  \includegraphics[width=6cm]{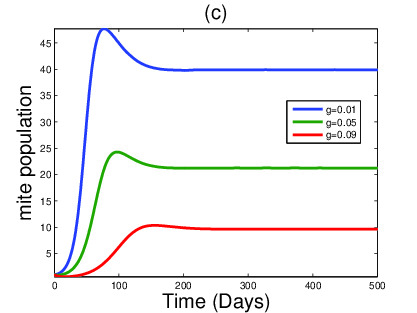}
\end{minipage}\hfil
\caption{\it Time-varying Plots of pupae cells, adult bees and mites with variations of the grooming rate of adult bees with initial conditions: $x(0)=7000,y(0)=15000,z(0)=1.$}
\end{figure}
\subsubsection{Keeping Varroa Sensitive Hygiene (VSH) Bees}
\paragraph\ The Varroa Sensitive Hygiene adult bees kill and reduce reproductive mites in mite-infested pupae cells. The VSH mite-control strategy is intended to limit mite reproduction. This mite-control strategy directly affects the hygienic rate parameter $h.$ The VSH adult bees kill and reduce reproductive mites in mite-infested pupae cells at a minimum rate of $8\%$ per day (Santos \emph{ et al.,} 2016). In this study, the effect of hygienic rate, $h$ on the pupae cells is investigated by varying it in the range of $4\%-12\%$ per day. Results of Figures $5(a)$ and $5(b)$ show that increasing values of $h$ increases the pupae cell and adult bee population. On the other hand, Figure $5(c)$ shows that increasing the values of $h$ reduces the mite population. From the third equation of Model $(1),$ the average life span, $\frac{1}{h+g+d_{m}}$ of a mite decreases as $h$ increases. This accelerates mite mortality leading to a reduction of the parasitism rate $(\alpha)$. The reduction in $\alpha$ increases the average life span, $\frac{1}{\sigma+m+\alpha z}$ of the pupae cells (Equation $(1)$ of Model $(1)$) which increases their number. Consequently, the adult bees increase leading to an increase in  colony performance.
\begin{figure}[H]
\begin{minipage}{0.3\textwidth}
  \centering
  \includegraphics[width=6cm]{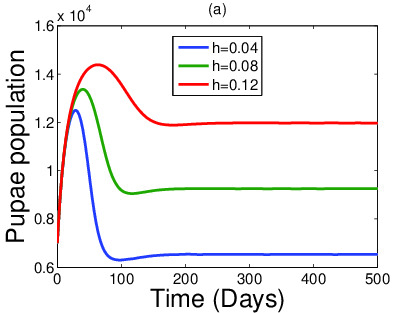}
\end{minipage}\hfil
\begin{minipage}{0.3\textwidth}
  \centering
  \includegraphics[width=6cm]{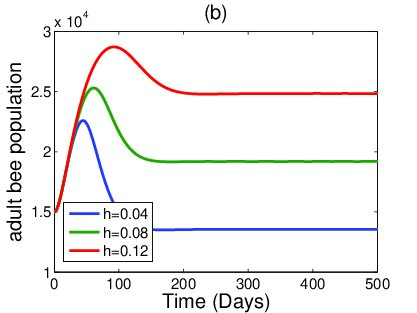}
\end{minipage}\hfil
\begin{minipage}{0.3\textwidth}
  \centering
  \includegraphics[width=6cm]{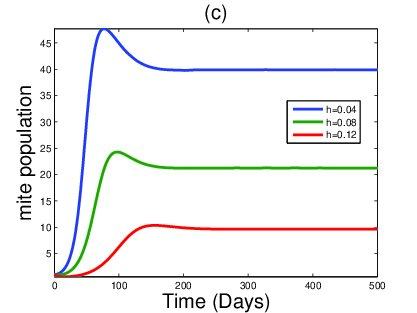}
\end{minipage}\hfil
\caption{\it Time-varying Plots of pupae cells, adult bees and mites with variations of the hygienic rate of adult bees with initial conditions: $x(0)=7000,y(0)=15000,z(0)=1.$}
\end{figure}
\subsubsection{Mite-trapping/Installing Screened Bottom Boards in Bee-hives/Spraying the Bees with Organic Chemicals.}
\paragraph\ The three mite-control methods above are applied to kill and reduce mites in the colony and therefore regulate parasitism rate $(\alpha).$  We simulate the effect of the three mite-control methods and observe the population dynamics of adult bees, pupae cells and mites. Existing studies have not quantified the parasitism effect by mites on honey bees. In this study, we consider and simulate the effect of the three control methods above by varying $\alpha$ in the range $0.003-0.007.$ Subsequently, results of Figures $6(a)$ and $6(b)$ show that a reduction in $\alpha$ leads to an increase in both the adult and pupae populations. This is because a reduction in $\alpha$ implies a reduction in the number of mites in the colony (Fig.$6(c)$). From the first equation of Model $(1),$ a decrease in the parasitism rate $\alpha$ increases the average life span, $\frac{1}{\sigma+m+\alpha z}$ of the pupae cells. This increases the number of pupae cells which in turn increases the adult bees thus, increasing colony performance.
\begin{figure}[H]
 \begin{minipage}{0.3\textwidth}
  \centering
  \includegraphics[width=5.7cm]{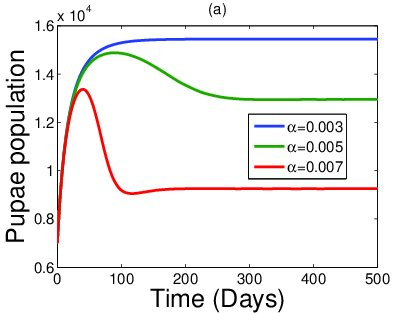}
 \end{minipage}\hfil
 \begin{minipage}{0.3\textwidth}
  \centering
  \includegraphics[width=5.7cm]{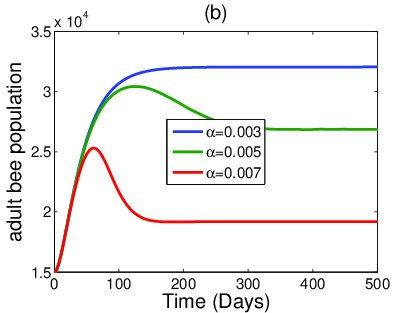}
 \end{minipage}\hfil
 \begin{minipage}{0.3\textwidth}
  \centering
  \includegraphics[width=5.7cm]{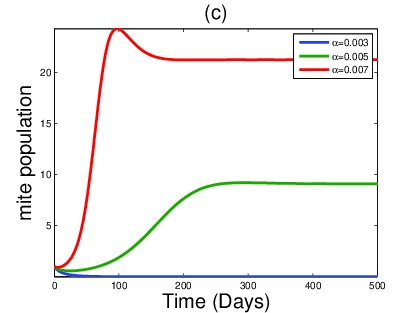}
 \end{minipage}\hfil
 \caption{\it Time-varying Plots of pupae cells, adult bees and mites with variations of the parasitism rate with $\epsilon=0.7,c=0.007$. Initial conditions are: $x(0)=7000,y(0)=15000,z(0)=1.$}
 \end{figure}
\section{Discussion of Results and Conclusion}
\paragraph\ In this study, a mathematical model for the population dynamics of honeybees infested with mites has been formulated and analysed. We have evaluated the effect of various mite-control methods on mite population growth and colony performance. For example; using a young queen bee, caging the queen bee, mite trapping using artificial drone combs, sprinkling bees with powdered sugar, keeping Varroa Sensitive Hygiene bees, spraying bees with organic chemicals and installing screened bottom boards in bee hives have been found to be vital mite control strategies in honey bee colonies. This study has been of great significance because previous mathematical models in the literature had given less attention to the evaluation of these mite-control methods.
\paragraph\ The long term behavior of the model has revealed various results. For example; a healthy colony naturally resists getting extinct when basic demographic reproduction number, $R_{h}>1$ otherwise it goes extinct when $R_{h}<1.$  The results have also shown that it is possible to eradicate mites from the colony if mite reproduction number, $R_{m}<1$ and $R_{h}>1$ otherwise the mites and honeybees coexist if $R_{m}>1$ and $R_{h}>1.$ This is in agreement with various mathematical studies that have modelled the interaction between honey bees and mites (Kang \emph{et al.,} 2016; Bernadi and Venturino, 2016; Denes and Mahmoud, 2019). However, this study gives an additional condition concerning the demographic basic reproduction number, $R_{h}>1$ that ensures colony survival. Sensitivity analysis suggests that introducing a young queen bee is one of the most effective mite control methods followed by caging the queen bee. Mite-trapping, adding screened-bottom boards, spraying the bees with organic chemicals are also reliable mite control strategies that are recommended by this study. In addition, sprinkling the bees with powdered sugar and keeping Varroa Sensitive Hygiene bees have a profound effect of lowering mite-infestation and hence improving colony performance. Therefore, bee-farmers are advised to always replace the aging queen bees with young ones in their colonies. This is because introducing a young queen bee, has a significant effect on the mite reproduction number, $R_{m}$ compared to other mite control strategies. Besides introducing a young queen bee, we recommend to the bee-farmers to always cage the queen bee at regular intervals (2-3 weeks) depending on the level of bee mite-infestation and on the number of brood cells in the colony. In this study, the numerical simulation results seem to agree with the analytical results that it is possible to control mites in the colony using intervention strategies aimed at improving colony performance.
\paragraph\ For effective mite-infestation control, a bee-keeper may opt to use a combination of more than one control strategy. For example, a bee-keeper may decide to keep Varroa Sensitive Hygiene bees (to kill reproductive mites inside the pupae cells) and at the same time sprinkle powdered sugar on the bees to induce grooming of mites off their bodies. African bees are known to be highly hygienic, efficient at mite grooming and have a shorter post capping period compared to European bees (Chemurot \emph{et al}., 2016; Kasangaki \emph{et al}., 2018). These three traits enable them to significantly reduce mite population in mite-infested colonies.\\
{\bf \large Acknowledgement}: The authors would like to thank Dr. Chemurot Moses who is a lecturer in the department of Zoology, Entomology and Fisheries Sciences, College of Natural Sciences of Makerere University for introducing them to Apiculture which is a versatile area of research. We would also like to thank the bee farmers of Luwero for exposing us to the practical aspects of bee keeping.\\

{\bf {\large References:}}
\begin{enumerate}
\item Aky$\ddot{o}$l, E., Yeninar, H., Karatepe, M., Karatepe, B. and \"{O}zk\"{o}k, D. ($2007$). Effects of queen ages on Varroa (Varroa destructor) infestation level in honey bee (Apis mellifera caucasica) colonies and colony performance. {\it Italian Journal of Animal Science,} $\bf{6}$:
    $143-149 $.
\item Bakare, E.A., Are, E.B., Abolarin, O.E., Osanyinlusi, S.A., Benitho, N. and Obiaderi, N.U. ($2020$). Mathematical Modelling and Analysis of Transmission Dynamics of Lassa Fever. {\it Journal of Applied Mathematics,} Article ID 6131708 18 pages: https://doi.org/10.1155/2020/6131708
\item Bernadi, S. and Venturino, E. ($2016$). Viral epidemiology of adult Apis mellifera infested by Varroa destructor mite. {\it Heliyon}. $\bf{2}$(5): $e00101.$
\item Boncristiani, H., Ellis, J., Bustamonte, T., Kimmel, C., Jack, C., Mortensen, A. and Schmehl, D. (2021). World honey bee health: the global distribution of western honey bee (Apis mellifera L.) pests and pathogens. \emph{Bee World,} $\bf{98}$(1): $2-6.$
\item Castillo-Chavez, C., Feng, W. and Huang, W. (2002). On the ccomputation of $R_{0}$ and its role on global stability, mathematical Approaches for Emerging and Reemerging Infectious Diseases: \emph{An introduction. The IMA Volumes in Mathematics and its Applications,} $\bf{125}$:
    $229-250$, Springer, New York.
\item Cameron, J. and James, D.E. ($2021$).  Integrated Pest Management Control of Varroa destructor(Acari: Varroidae), the Most Damaging Pest of (Apis mellifera L. (Hymenoptera: Apidae)) colonies. {\it Journal of Insect Science}, $\bf{21}$(5): $1-32.$
\item Chemurot, M., Akol, A.M, Masembe, C., Lina de Smet, Descamps, T. and Dirk C. de Graaf.  ($2016$).  Factors influencing the prevalence and infestation levels of Varroa destructor in honey bee colonies in two highland agro-ecological zones of Uganda. Exp Appl Acarol DOI10.1007/s10493-016-001-x
\item Chien, F. and Shateyi, S. ($2021$).  Volterra-Lyapunov Stability Analysis of the Solutions of Babesiosis Disease Model. {\it Symmetry}, $\bf{13}$: $1272.$
\item Denes, A. and Mahmoud, A.I. (2019). Global dynamics of a mathematical honeybee colony infested by virus-carrying Varroamites. {\it Journal of Applied Mathematics and Computing,} $\bf {61}$: $349-371.$
\item Di Prisco, G., Annoscia, D., Margiotta, M., Ferrara, R., Varricchio, P., Zanni, V. (2016). A mutualistic symbiosis between a parasitic mite and a pathogenic virus undermines honey bee immunity and health. \emph{Proceedings of the National Academy of Sciences}, $\bf{113}$: $3203-3208.$
\item Fries, I., Aarhus, A., Hansen, H. and Korpela, S.  ($1991$). Comparison of diagnostic methods for detection of low infestation levels of Varroa jacobsoni in honey bee (Apis mellifera) colonies. {\it Experimental and Applied Acarology}, \textbf{10}(1991): $279-287.$
\item Harbo, R.J. ($1986$). Effect of population size on brood production, worker survival and honey gain in colonies of honeybees. {\it Journal of Apicultural Research,} $\bf{25}$(1): $22-29.$
\item Kang, Y., Blanco, T., Wang, D.Y. and Degrandi-Hoffman. ($2016$). Disease dynamics of honeybees with varroa destructor as parasite and virus vector. {\it Mathematical Biosciences}, $\bf{275}$: $71-92.$
\item Kasangaki, P., Nyamasyo, G., Ndegwa, P. and Kajobe, R. ($2018$). Assessment of Honeybee Colony Performance in the Ecological Zones in Uganda. {\it Current Investigations in Agriculture and Current Research,} $\bf{1}$(5): CIACR.MS.ID.000121.
\item Khoury, D.S., Myercough, M.R. and Barron, A.B. ($2011$). A quantitative model of honey bee colony population dynamics. \emph{Public Library of Sience One,} \textbf{6}(4): e18491.
\item Li, M.Y. and Muldowney, J.S. ($1995$). A geometric approach to global stability problems. \emph{Siam Journal of Mathematical Analysis}, \textbf{27}: $1070-1083.$
\item Liao, S. and Wang, J. ($2012$). Global stability analysis of epidemiological models based on Volterra-Lyapunov stable matrices. {\it Solitons and Fractals}, $\bf{45}$: $966-977$.
\item MacCluskey, C.C. and van den Driesche, P. ($2004$). Global analysis of tuberculosis Models.
     \emph{Journal of Differential Equations,} \textbf{16}: $139-166$
\item Marco P. and Giovanni F. ($2018$). Liquid formic acid $60\%$ to control varroa mites (Varroa destructor) in honey bee colonies (apis mellifera: protocol evaluation): {\it Journal of Apicultural Research}, $\bf{57}$(2): 300-307.

\item Messan, K., DeGrandi-Hoffman, G., Castillo-Chavez, C. and Kang, Y. ($2012$). {Migration Effects on Population Dynamics of the Honeybee-mite interactions.} Maths.Model. \emph{Mathematical Modelling of Natural Phenomena} \textbf{7}(2): 32-61.
%\item Li, M.Y. and Muldowney, J.S. ($1995$). A geometric approach to global stability problems. \emph{Siam J. Math. Anal}, \textbf{27}: 1070-1083.
%\item Ratti, V., Kevan, P.G. and Eberl, H.J. (2013). A mathematical model for population dynamics in honeybee colonies infested with Varroa destructor and Acute Bee Paralysis Virus. \emph{Canadian Applied Mathematical Quarterly} \textbf{21}(1):$ 63-93.$
%\item Russel, S., Barron, A.B. and Harris, D.(2013). Dynamic modelling of honey bee (Api mellifera) colony growth and failure. \emph{Ecological Modelling} \textbf{265}: $ 158-169.$
\item Santos, J.F., Coelho, P.J. and Blimann, P.J.  ($2016$).  Behavioral modulation of the coexistence between Apis mellifera and Varroa destructor: a defense against colony collapse. \emph{PeerJ Pre-Prints,} e1739.
\item Schmickl, T. and Karsai, I. (2016). How regulation based on a common stomach leads to economic optimization of honeybee foraging. \emph{Journal of Theoretical Biology,} \textbf{389}:  274-286.
\item Sumpter, D.J.T. and Martin, S.J. (2004). The dynamics of virus epidemics in varroa-infested honeybee colonies. \emph{Journal of Animal Ecology} \textbf{73}: 51-63.
\item Thieme, H.R. ($2003$). Mathematics in population biology. Princeton University Press.
\item Tian, P.V. and Wang, J. (2011). Global stability                  for cholera epidemic models. \emph{Mathematical Biosciences} \textbf{232}(1): $31-41.$
\item Torres, J.D. and Torres, A.N. (2020). Modelling the Influence of Mites on Honey Bee Populations. \emph{Veterinary Sciences,} \textbf{7}(3): $139$.
\item Van den Driessche, P. and Watmough J. (2002). Reproduction numbers and sub-threshold endemic equilibria for compartmental models of disease transmission. \emph{Mathematical Biosciences,} $\textbf{180}$: $29-48$.
\item Ziyad, A., Atif, I., Rashid, M., Ghulam, S., Muhammad, A., Saboor, A., Muhammad, M.R. and Jun, L. ($2021$). {Effectiveness of Different Soft Acaricides against Honey Bee Ectoparasitic Mite Varroa destructor (Acari: Varroidae).} \emph{Insects} \textbf{12}(11): 1032.
\end{enumerate}
\end{document}